# Analysis of the Main Factors Affecting M-Commerce Adoption in Iraq.

Alaa Mahdi Sahi

College of Administration and Economics, Wasit University

**ABSTRACT**

The telecommunications sector in Iraq is one of the most dynamic and active industries. This sector has prime importance only next to oil & gas, as revealed by recent reports. This industry has grown in leaps and bounds with the rising demand for mobile services. Mobile services were inaccessible prior to 2003. As of 2016, there are 29 million mobile telephone subscribers, with only 2.3 million fixed-line users. With the easy availability of telephone services, the telecommunication sector has reported a rapid boom. Subsequent to the fall of Saddam Hussein's regime on April 9, 2003, mobile commerce has reached new heights. This phenomenon is mainly attributed to the increasing internet and smartphone usage. Telecommunication is now an imperative element in the social and economic growth of the country. This sector has also expanded access to various prospects and has altered the manner of consumer interaction, comparing prices, researching goods, and making purchases. Several studies have reviewed the penetration of mobile commerce to varied sectors. These studies have also delved on the challenges, prospects, concerns, and impediments in this sector. However, none of these studies has considered the situation in Iraq. This paper examines the current literature to ascertain the parameters affecting m-commerce adoption in Iraq

**KEYWORDS:** Mobile Commerce, Smartphone Penetration, Mobile Internet, Adoption & Iraq

## II. INTRODUCTION

Mobile commerce (m-commerce) has revolutionized the manner in which technologies can alter business pattern of consumers engaged in electronic business. Consumers anticipated it to involve mobile "electronic commerce." They reckoned that it would allow them to buy goods and services with their wireless mobile devices everywhere, at all times. This mobility, sustained by mobile telecommunications network, is the defining factor that distinguishes mobile computing from other information technology applications. Globally, with increase in digital economy, m-commerce has become a vital element in business strategy. In addition, its usage has grown over the recent years, making it the most significant development in business. With increasing mobile penetration, mobile commerce has also risen to unimaginable heights. In web Retailer's ((2014)) Mobile ((500)) study ((2013)), mobile sales for its ((500)) biggest sellers Across the Globe escalated by 71% over the same time frame in 2012, reaching $30.5 billion (up from $17.8 billion) [1].In Iraq, cell phone/mobile phone communication has been established only in this decade, and only in the Kurdistan Region. In the initial times, there was only one tower, which resulted in minimal coverage In addition, a line was very expensive ($400)[2]. The progress, thereafter, has been massive. Currently, four main cell phone companies operate in Iraq, which are Asia cell Communications, Zain Iraq (formerly MTC Atheer), Korek Telecom, and Regional Telecom/ Fast link. Central statistics organization-Iraq conducted a survey revealing that the year 2014 saw most families having access to a mobile phone (prepaid card) (98.88%). Either in urban (99.13%) Versus (98.15%) in rural areas compared to 2008 where she was for (94.1%) Across Iraq, urban (96.1%) Versus (90.0%) In rural areas the proportion of monthly subscription (Bill) (0.57%) In urban (0.70%) Versus (0.18%) In rural areas compared to 2008 (0.1%) [3]

### *Research Problem*
In the urban areas of Iraq, internet and mobile phone usage is widespread among the younger generation. However, the youth are not exposed to e-commerce and m-commerce. This paper emphasizes on the factors affecting m- commerce adoption in Iraq.

### *Paper Question*
What are the foremost elements affecting mobile commerce adoption in Iraq?

### *Goal of this Paper*
To ascertain the factors affecting mobile commerce adoption in Iraq.

### *Purpose of Study*
- Investigate the crucial features that are an obstacle M commerce ((m commerce through cellular phone or some other wireeless device)) adoption , in Iraq
- Study the Overview of m commerce.
- Understand the association between m commerce & e commerce.
- Deliberate on the current benefits and demerits of m-commerce.
- Characterize the m-Commerce applications.

### *Importance of the Study*

This paper intent to assist the local businesses have a greater perception regarding the obstacles of mobile commerce and the intention of the consumers to adopt this feature in Iraq.

**II.LITERATURE REVIEW**

Definitions of Mobile Commerce M-commerce is also known as "mobile e-commerce" because all its transactions are electronic in nature, with a mobile terminal and a wireless network. Mobile terminals comprise portable devices including mobile telephones, PDAs, and devices "mounted in the vehicles that are capable of accessing wireless networks." Any of these devices should be capable to perform m-commerce transactions[4]. M-commerce utilizes wireless technologies to improve the reach of e-commerce applications. Transactions with mobile applications can be by the customer or supplier and helps to conduct e- business both within and across an organization's boundaries. As a result of its varied roles, it is considered as an integral part of any organization's strategy[5]. M-commerce is defined as any direct or indirect transaction managed and expedited through a wireless telecommunication network. It involves the usage, application, and amalgamation of wireless telecommunication technologies and wireless devices within the business systems domain. The term m-commerce incorporates reference to the infrastructures and electronic technologies required for wireless data and information transfer, along with all its multimedia forms (i.e. text, graphics, video, and voice)[6]. UNCTAD's E-Commerce and Development Report 2002 defines "M-commerce as the buying and selling of goods and services, using wireless hand-held devices such as mobile telephones or personal data assistants (PDAs)"[7]

*Difference between M-Commerce and E-Commerce*

Electronic commerce or e-commerce refers to any electronic buying and selling on the Internet [8]. Accordingly, Mobile-commerce is recognized as an expansion of electronic-commerce, because the two terms are more or less alike. Their essential business principles are similar [9]. According to Rainer (2000), m-commerce pertains to the use of wireless communications technology to obtain network-based information and applications with mobile devices [10]. However, m-commerce has some elements that make it distinctive from all other forms of electronic commerce.

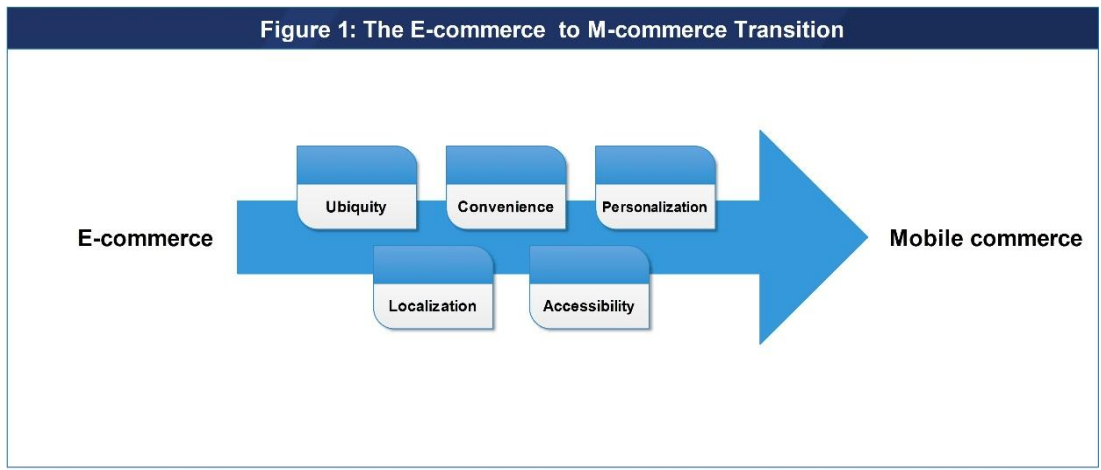

Figure 1: The E-commerce to M-commerce Transition

Source: Junglas, I. and R. Watson, U-commerce: a conceptual extension of e-commerce and m-commerce. ICIS 2003 Proceedings, 2003: p. 55.

- **Ubiquity:** Ubiquity is the main benefit of mobile commerce. Users can access any interested information at any location. They do so using Internet-enabled mobile devices. [11]
- **Convenience:** Convenience is associated with generating agility and easy accessibility. This in delivered by developing user-friendly wireless handheld devices, which are not dependent on time or location [10].
- **Localization:** Mobile operators or positioning technology can help track the location of any mobile device. This aspect can be researched to develop unprecedented opportunities in innovative location-based services [12].
- **Personalization:** The Internet is crammed with information, services, and applications. However, different personalized services and applications should be build according to the preferences of the individual mobile user[13].
- **Accessibility:** Contrary to m-reachability, m-accessibility deals with a user's access to the mobile network at any time and from any location. This can only be possible with ample mobile network coverage.[14].

*Advantages of Mobile Commerce*

Mobile commerce improves value-added utility in the transactions of users, especially in the following situations [15]

- **Context-Specific Services:** Mobile commerce can profit from location-based services. These services may be restricted to a particular context (e.g. time of the day, location of the user, and individual interests). These services create new opportunities for personalized push-marketing. This type of commerce is in close proximity to the vendor, so there is a direct increase in sales. It encourages brand presence, and consumers tend to rely on brands they are familiar with.
- **Spontaneous Decisions and Needs:** Spontaneous needs are not influenced by the external market but are decisions generally taken at the spur of the moment. These decisions involve only small sums of money. An example of such a service is booking a place in a restaurant or cinema spontaneously. Users may also spontaneously use entertainment facilities, e.g. horoscope, music, or sport news, while travelling and when free.

- **Efficiency Increase:** Mobile commerce is reported to increase the efficiency of the workforce. Employees (consumers) who are time constrained can avail 'dead spots' in the day, e.g. while travelling to and from the workplace, to make their purchases, check E-mails, get current news, and complete bank transactions.

**Disadvantages of Mobile Commerce**
- A major drawback is the mobile device, as it may not be able to provide the same level of graphics or processing power of a personal computer. The mobile device is constrained by the technology employed to built it[16].
- The size of the screen in a mobile device may limit access to complex of applications[17].
- Mobile electronic commerce becomes inconsistent because of non-uniformity of networks. Each network may approach mobile commerce differently[18].
- M-commerce may be limited by the network service, as they need to depend on its reliability to perform any transaction[19]. Mobile devices have been reported to lose connectivity or have a short battery life. This limits the time the consumers have to complete their business without losing connectivity mid-transmission .The security of mobile electronic commerce is another major concern[19].

## IV. AN M-Commerce Consumer-Centric Model

Coursaris and Hassane in from McMaster University in Canada postulate a model that assists in defining value proposition to m-consumers. First, the m-consumer interaction modes within a wireless environment are identified[20]. By identifying the m-consumer's possible activities, the following entities were identified. These entities help to understand the interaction required, and to what degree:

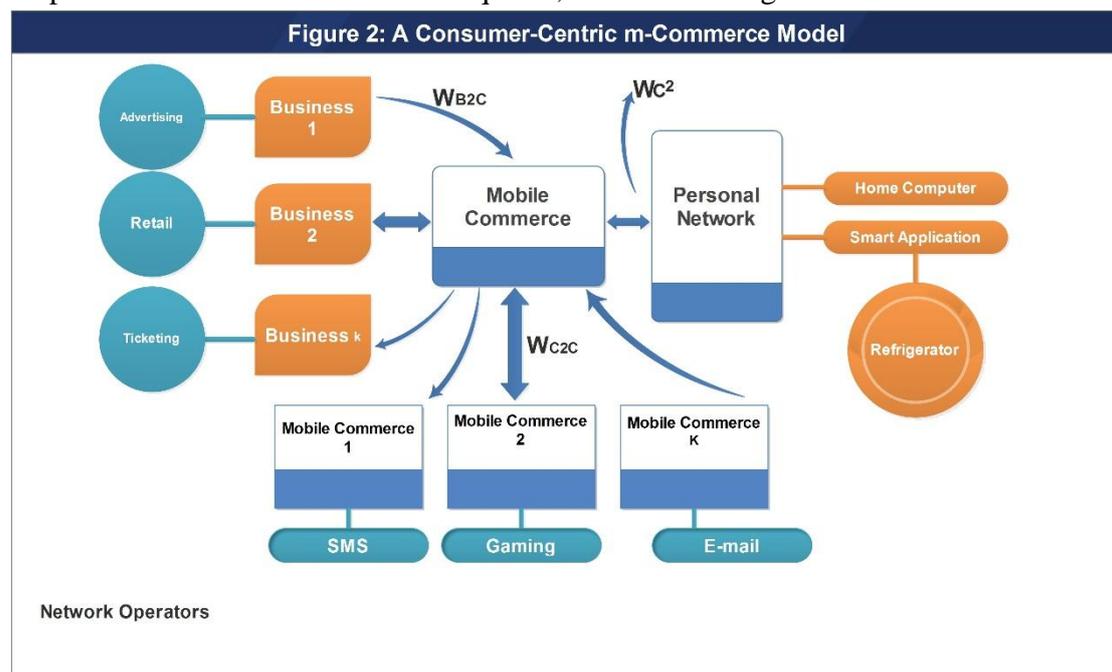

Source : Coursaris, C. and K. Hassanein, Understanding m-commerce: a consumer-centric model. Quarterly journal of electronic commerce, 2002. 3: p. 247-272.

- ***Businesses:*** Involving a Wireless Business-to-Consumer (WB2C) interaction mode. Most interactions operate using this mode, as well as a Wireless Consumer-to-Business (WC2B) interaction mode.
- ***Consumers:*** Involving a Wireless Consumer-to-Consumer (WC2C) mode of interaction.
- ***Personal Networks:*** Involving a Wireless Consumer-to-Self (WC2) interaction mode.
- **Businesses:** Refer to individuals or organizations that a consumer has to interact with wirelessly for business- related purposes. In addition, consumers are generally at the receiving end of any interaction initiated by businesses [21]. In this paper, WB2C represents such interactions, without stipulating which party initiated the interaction [21]. Some examples of such a business interaction Comprise advertising and retail offers targeted at m consumers [21].

- ***M Consumers:*** Are those people who may desire to interact wirelessly for private functions. It is regarded as WC2C form of interaction. It has communications (for example E mail & SMS) and entertainment ( for instance gaming in a multiplayer format) [21].

- **Personal Network:** Is the server that a shopper possesses and the personal network may be accessed wirelessly for personal purposes. This type of interaction is denoted as (WC2) [21]. Cases of this type of interaction comprise a mobile user participated in wireless communications Using In-house House computer and its System, in Addition to Some Other Intelligent appliances, which Might be connected to that System (for instance a fridge) [21].

- ***Mobile-Based Services:*** Mobile devices can provide various remote services such as m-commerce, m-health, m- education and m-Government [22]. Mobile devices are the most easily available ICT devices. Its technology has proved so successful that New cell phone devices offer highspeed communication channel, with 4G and 3G broadband networks and brag strong core processors [22]. Hence, prices for devices has drastically reduced and usage of communication channels has gotten better. In 2014, the usage of cell phone device has attained nearly all the people in the world [22].

Second, mobile devices allow user connectivity at anytime and anyplace, even while travelling. Several organizations have initiated research to harness the potentials of mobile devises. This also includes the The International Telecommunication Union, which has established standards for the security of cellphone financial transactions. The ITU shares its experience in the field of fresh developments. Throughout 2010 / 2014, Study Group 2 of the ITU Telecommunication Development Sector (ITU-D) developed a "Toolkit for creating ICT-based services employing mobile communications for E government services." This toolkit was meant to explain the using of those standards and to demonstrate the building of mobile based systems across countries.

**V. Classification of M Commerce Services**

M commerce integrates Mobile Financial Services (MFS) and m Marketing. As observed in Fig 2, MFS includes mBanking, mPayment and mMoney transfer. Therefore, mMarketing is a collaboration of mobile informing, mobile loyalty and mobile remote capture.

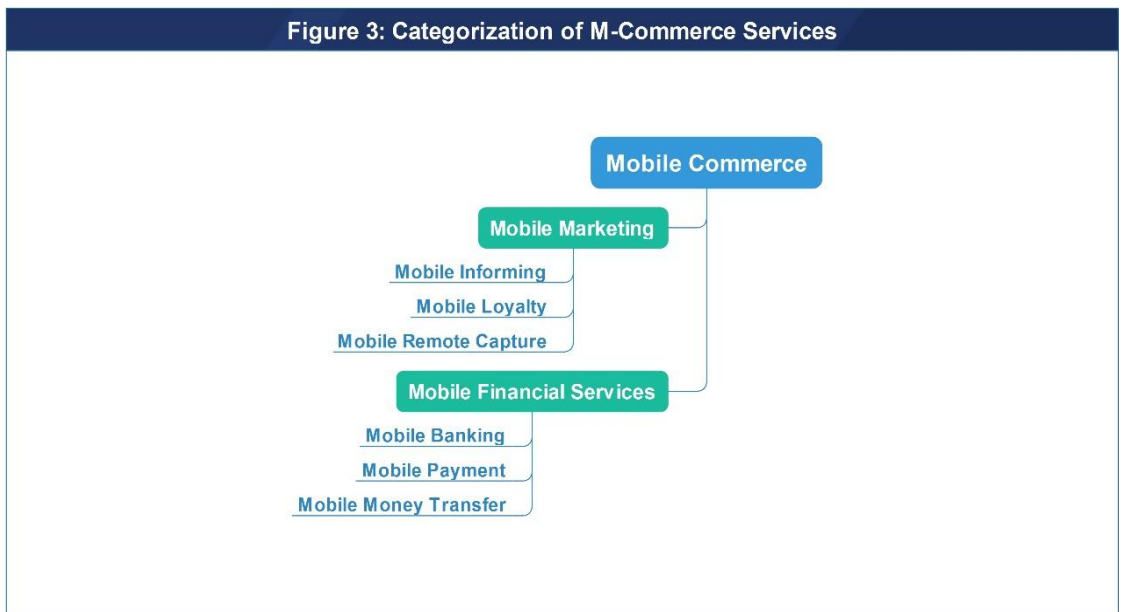

Source: International Telecommunication Union

- **Mobile financial services(MFS):** It involves the usage of mobile phone while managing financial services and performing financial transactions. It is a combination of both transactional and non-transactional facilities, such as viewing financial information on one's mobile phone[23].
- ❖ **M-Banking:** This term is often used to illustrate banking business performed using mobile devices such as mobile phones or PDAs[24]. • M-Payments: M-payments includes payments for products or services rendered between two parties, wherein a mobile device, such as a mobile phone, is the primary object used for completion of the work[25].

- ❖ ***Mobile Money Transfer:*** Mobile money transfer included all services in which the customers utilize their mobile devices to either send or receive money. It also includes the electronic transfer of money from one individual to another using a mobile phone. This application embraces both domestic transfer or international remittance transaction[26].
- ***M-Marketing:*** Mobile Banking is described as performing banking business using mobile devices such as mobile phones or PDAs[27]. This process involves the following three approaches:
- ❖ ***Mobile Informing:*** Mobile Informing is an data service that employs cellphone devices. Mobile informing is most advantageous because the information is sent directly into the consumer´s device. The information may be made available anywhere, at any time, through cellular networks. For instances of such services comprise bank informing and advertisement [28].
- ❖ ***Mobile Loyalty:*** Mobile Loyalty is a loyalty system based on mobile devices.
- ❖ ***Remote deposit capture (RDC):*** is a technique that permits a customer to scan checks remotely and transmit the images of the check to a bank for deposit. This procedure usually involves an encrypted Internet connection[29].

## VI. M-Commerce Applications

The following are the most popular m-commerce applications:

- **Mobile Ticketing**

Mobile ticketing is one among the many m-commerce applications provided by wireless network operators. To procure a cinema ticket, for example, the user can avail an electronic receipt, which is considered as proof of purchase for the cinema ticket[30]. However, the electronic receipt particulars must meet a list of requirements, such as it should be acknowledged by the delivery entity;[30] it should be valid as proof for the holder of the receipt that a purchase has been made and goods can be delivered; it should not be falsified and incapable of duplication or repeated usage[30].

- **Mobile Vouchers, Coupons and Loyalty Cards**

Mobile ticketing technology also involves the distribution of vouchers, coupons and loyalty cards[31]. The voucher, coupon, or loyalty card sent to the mobile phone signifies a virtual token. A consumer can present a mobile phone with one of these tokens at the point of sale and can avail similar privileges as another customer who has a loyalty card or other paper coupon/voucher[31]. Mobile vouchers allow merchants close the loop on mobile marketing—starting when the customers first see an ad and ending with redemption at the point of sale[32]. With significant data, marketers can more efficiently stimulate consumer patterns and calculate the returns on a campaign more precisely[32]. Mobile vouchers enable mobile marketing campaigns to be more measurable, while being a launch pad for varied and flexible campaigns[32].

- **Mobile Banking**

M-banking involves all customer interactions with a bank via a mobile device, such as a mobile phone or personal digital assistant (PDA)[15]. The exclusively involves data communication, and so does not include telephone banking in any form, either in its traditional form of voice dial-up or by the dial-up to a service based on touch tone phones[15].

- **Mobile Brokerage**

In banking and financial services, brokerage includes all intermediary services pertaining to the bourse, e.g. selling and purchasing of stocks[32]. Mobile Brokerage is therefore transaction based, mobile financial services of non-informational character that are related to a securities account[32].

- **Mobile Brokerage**

may be segregated in two categories to distinguish between services that are mandatory to operate a securities account and services that are mandatory to administer that account[32].

- **Mobile Purchase**

Mobile purchase permits customers to shop online anytime, anywhere. Customers can browse and purchase using any cheap, secure payment mode[33]. In place of paper catalogues, retailers can send customers a products' list that entails to the customer's need. This can be either directly via a mobile device or the consumer can visit a mobile version of the retailer's e-commerce site[33].

- **Mobile Marketing and Advertising**

Mobile marketing is a two-way or multi-way communication. It involves advertising an offer between a firm and its customers using a mobile medium, device, or technology[20]. It is an interactive network, and could include mobile advertising, promotion[20], customer support, and other relationship-building activities. Such collaborating marketing activities are gaining prime importance with the evolving business landscapes[20].

**VII. Mobile Payment**

Mobile payment services have a key role to play in mobile commerce. Mobile payment may be a stand-alone service[34], but it has huge potential when incorporated with other m-commerce services. To gain user acceptance, these services must be made more secure and user-friendly[34]. The history of m payment solutions is brief; however, the sector has shown rapid improvement [35]. The Finnish telecom operator Sonera in 1997 opened one of the most important and first portable payment solution services. Their service allowed the purchase of goods via vending machines with cellphones and payment through cellphone operator [35]' service invoices along with the mobile services. New inventions used in m payment solutions and new application fields for m payments have been established at an increasingly rapid pace ever since [35]. Mobile payment (mPayment) involves payments using mobile devices, such as wireless handsets, personal digital assistants (PDA), radio frequency (RF) devices, and near field communication (NFC) based devices. The payment industry is anticipating that mPayment will deliver the convenience, transaction speed, and versatility essential for today's complex marketplace. Although mPayment is still in its initial stages, experts forecasta $37 billion increase in global mPayment transaction volume by 2008[36].

## VIII. MOBILE MARKET IN IRAQ

### *COUNTRY OVERVIEW*

Violence persists in the country, and this has a major impact on telecommunications development. There have only brief respites from the political tensions within the country over the past ten or more years. Even in 2016, problems between the existing government and sectarian groups continue. Rebel forces have dominated parts of Iraq. Iraq's economy is centralized on the oil sector, which accounts for over 40% of GDP and 90% of government revenue. Oil production has progressively increased with investments in infrastructure. In spite of this, constrains in pipeline capacity and the security situation poses challenges. The International Energy Agency estimates that Iraq has the capacity to earn up to US$5 trillion from oil exports by 2035, or US$200 billion per year.The Iraqi Government's National Development Plans aims at producing a diversified economy that is not just dependent on oil. One of the main objectives of the National Development Plan is to encourage private Iraqi capital investment in Iraq, as substantial amounts are currently invested outside Iraq.

**Table 1 – GDP growth and inflation – 2006 - 2016**

| Year | Real GDP Growth (% change yoy) | Inflation (% change yoy) |
|------|-------------------------------|--------------------------|
| 2010 | 6.4% | 2.4% |
| 2011 | 7.5% | 5.6% |
| 2012 | 13.9% | 6.1% |
| 2013 | 6.6% | 1.9% |
| 2014 | -2.1% | 2.2% |
| 2015 | 0.0% | 1.9% |
| 2016 | 7.1% | 2.0% |

**Sources:** IMF World Economic Outlook Database, DFAT

The rate of unemployment varies in the data from different sources. However, the average is considered to be between 12% and 23% . Corruption is the major hinderance to economic development. Iraq's Commission of Integrity, in association with the Central Statistics Office and United Nations Development Programme (UNDP) Iraq, conducted a survey in 2013 that revealed that over 31,000 civil servants Considered corruption as increasing. In comparison to other countries, Iraq rates poorly, as is manifested by the Transparency International's Corruption Perception Index for 2013, which ranked Iraq at 171 out of the 177 countries surveyed. In 2015, the ranking of Iraq was relatively the same with a ranking of 161 out of 168 countries surveyed. In 2016, Iraq's economy had undue pressure from the expenses of financing war as well as the humanitarian expenses for the millions of displaced and destitute people.

*Short History of Mobile Communication in Iraq*

The second Gulf war in 2003 further destroyed the already poor Iraqi communications infrastructure during its first week. Even before the war, Iraq had one of the weakest telecommunications services in the globe, primarily because of its destroyed economy, economic sanctions and the 1991 Gulf War . Before the 2003 war, Iraq had fewer than a million fixed - line subscribers and no national mobile network [37]. A healthy regulatory environment is yet to be founded in Iraq. In March 2004, the Communications Minister affirmed that a new telecoms and media law would be implemented soon in the future, aiming the privatization of the fixed-line network. However, it did not materialize. The US-led Coalition Provisional Authority (CPA) Order 11 approved the Ministry of Transportation and Communications as the licensing and regulatory body for all commercial telecommunication services in Iraq. The order was withdrawn after the transfer of full governance to the Iraqi Interim Government on 30 June 2004. The Ministry of Communications (MOC) was substituted by the Ministry of Transportation and Communications. The National Communications and Media Commission (CMC), which is independent of the MOC, was also instituted in March 2004 as a telecoms regulatory authority. A new Communications Law has been in draft for some time, but its authorization has been delayed.

## Mobile Network Operators of Iraq

Iraq's mobile operators are considered as star performers in the telecoms sector, in recognition of the speed in setting up their networks. The market grew swiftly, partially because of the absence of alternative telecoms services.

**Table 2: Major Mobile Operators in Iraq**

| |
|---|
| Asiacell Telecommunications Company Ltd |
| Zain Iraq |
| Korek Telecom Ltd |

Source: "ERICSSON REGION MIDDLE EAST COUNTRY REPORT: IRAQ". http://www.ericsson.com/. N.p., 2016. Web. 4 Dec. 2016.

- **Zain Group:** Zain is the foremost mobile and data services operator in 8 Middle Eastern and African countries. They have employed over 7,000 personnel to deliver far-reaching mobile voice and data services. As of June 30, 2016, they were reported to have over 45.2 million active individual and business customers [38]. The firm was established in 1983 in Kuwait as MTC or Mobile Telecommunications Company, and was later rebranded as Zain in 2007[39].
- **AsiaCell:** AsiaCell Telecom Company is an Iraqi telecommunications company that primarily focuses on providing mobile phone services and Mobile Internet in Iraq[39]. Asiacell was given a two-year GSM license for the six northern provinces of Iraq, tailoring to the needs of a wider client base who prioritized a quality mobile network[39]. The license was extended in 2005 to cover the entire Iraqi Republic. Currently, it is the only telecom network having nationwide coverage. Its GSM is 900 MHz (GPRS, EDGE)[39].
- **Korek Telecom:** Korek is a shared limited company registered in Iraq. Its focus is mainly in providing GSM services. Korek started its operations in Iraq in 2000. Its initial coverage area was the north of Iraq. It is the oldest Iraqi Telecom company[40]. On August 17th, 2007, Korek was awarded a nationwide mobile license. Currently, Korek has more than 32 million customers[40].

## Market Overview

Prior to the rise of violence, which continued in 2016, there were several positive developments in the country. The telecoms sector concentrated on improved network access. A number of wireless local loop and mobile licenses were issued with the prime focus on rebuilding damaged infrastructure. Numerous Internet cafes were commenced, which contributed to improved Internet access. For greater overall connectivity, several fixed-wireless licenses were awarded, which ensured the setting up of CDMA and WiMAX technology. Even the recent efforts concentrated on fibre, with the main focus on large-scale deployments to connect up to 2 million households. This effort was stalled by the recent conflict. International Internet connectivity was also improving, with additional capacity coming online.

The three mobile network operators that have national licenses are Zain Iraq, Asiacell and Korek Telecom. As per their pact, they needed to launch Initial Public Offerings (IPO)s as part of their license conditions by August 2011. Asiacell successfully launched its IPO in 2013, and in mid-2015, Zain Iraq followed suit.

In 2015, 3G services were launched by the three major mobile operators – Zain Iraq, Asiacell and Korek Telecom. The roll out of 3G was for a great leap in Iraq's development – this was offset by the iniation of new taxes. This increased the prices for Internet and mobile top-up cards by 20%. This proposal was part of the austerity measures by the Iraq government. The Iraqi Commission for Media and Communications (CMC) has suggested that a 4G LTE license allocation proposal would be put forward to the government in the near future. Prior to the ongoing civil tension, 4G LTE had been deployed in a few provinces of Iraq by Fastlink (Regional Telecom); in addition, fibre optic was also gradually being deployed across parts of Iraq. With greater access to mobile communication in Iraq, the fixed telecom sector is experiencing a steady decline

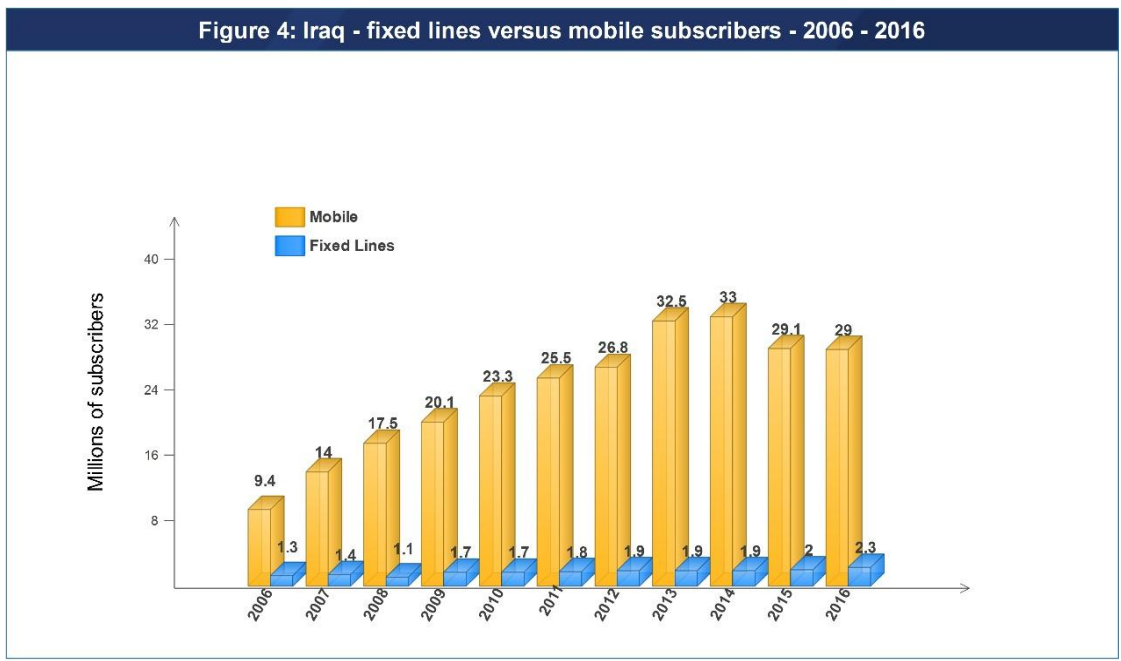

Source : Budde Comm Based on ITU 2016

The continuing civil war has resulted in several obstacles for mobile phone companies in Iraq, primarily because of the destruction of infrastructure, fluctuating subscriber base and revenue declines, all of which influence operations. Fibre optic cables were in place across several parts of the country; however, there are concerns that many of these would have been destroyed and will need to be rebuilt.

## *Market Analysis*

In Iraq, the mobile market was born post the United States invasion in 2003. Prior to that, there were no mobile operators in Iraq. Iraq is had access only to wire telephone. After the war in 2003 on Iraq and after the fall of ba'ath regime, restrictions were removed, and the mobile market became open and Telecom operators flooded Iraq. Iraq's mobile operators are acknowledged as star performers in the telecoms sector, because they set up their networks at great speeds. The market grew swiftly, partially because of the absence of alternative telecoms services[37]. 3 mobile phone licenses were granted at August 2007. The licenses replaced the earlier awarded temporary licenses to construct and operate 3 regional cellular networks covering the central northern and southern parts of Iraq [37]. The 3 operators extended their services and Realized Exactly the terms of the contracts. The 2007 license resulted in profound alterations in the cellphone market, as Oras com Telecom of Egypt did n't win a permit for its subsidiary Iraqna. has included not only the three licensed national operators but also an equal or more small regional operators. The Kurdish region north of Iraq has one very huge operator, Zain Iraq (which bought Iraqna), one large operator, As iacell (now with Ooredoo of Qatar as a major shareholder) and the two Kurdish operators – the larger of which, Korek Telecom, has a national license. The three main mobile network operators were required as per their respective licenses to be listed on the local stock exchange by 31st August 2011. However, the operators failed to meet the deadline, and talks are on with the government to resolve the issue. As iacell successfully launched its IPO in 2013 and Zain Iraq in mid-2015. The years 2014 and 2015 were one of the most challenging for Iraq, with escalating conflicts that resulted in massive relocation of civilians. This resulted is several mobile network zones becoming unreachable for repairs. Because of this, Iraq's mobile operators suffered stagnation or decline in subscribers and revenue. In late 2015 a fourth mobile operator license was to be issued, and the regulator called for tenders by December 10th. Mobile data services have the potential to generate new revenue given the maturing mobile voice market. In 2016, operators were initiating several new services and tariffs to improve the 3G service uptake. Table 2 illustrates the number of mobile phone subscribers in Iraq.

**Table 3 – Mobile subscribers and penetration rate – 2002 - 2016**

| Year | Subscribers (million) | Penetration |
|---|---|---|
| 2002 | 0.02 | <1% |
| 2003 | 0.08 | <1% |
| 2004 | 0.57 | 2.2% |
| 2005 | 1.53 | 5.6% |
| 2006 | 9.35 | 33% |
| 2007 | 14.02 | 49% |
| 2008 | 17.53 | 60% |
| 2009 | 20.12 | 67% |
| 2010 | 23.26 | 75% |
| 2011 | 25.52 | 80% |
| 2012 | 26.76 | 82% |
| 2013 | 32.45 | 96% |
| 2014 | 33.00 | 94% |
| 2015 | 29.10 | 81% |
| 2016 | 29.00 | 81% |

Source: BuddeComm based on industry data

### IX. Mobile Content and Applications

In Iraq, the progress in mobile content and applications usage is slow, primarily because of the lack of mobile broadband networks. The network operators had concentrated on SMS and MMS delivery based content, including access to email accounts and social media websites such as Facebook.

*M-Banking and M-Payments*

As iacell was the first operator to offer mobile payment services through the Mobicash platform. It permitted end users to transfer funds to other end users or between accounts of the same user. In 2016, As iacell introduced more advanced mobile money services through Ooredoo's Asia Hawala service, which permits funds

Impact Factor (JCC): 4.7129                                     NAAS Rating: 3.27

transfers, bill payments and airtime top-ups. Zain Iraq is also launching Zain Pay in phases in Iraq. Carrier billing services have already been rolled out in 2016.

*APPS*

Zain's mobile content and applications are offered by Zain Zone Iraq branded app. The portal permits users to access and download news, entertainment, music and video content. Zain's data services include 'Zain Create', which permits customers to buy and download music and videos. Zain provides credit transfer that allows users to transfer credit from one Zain account to another. Zain's ringback tone product is offered under the Sami3ny brand in Zain Zone.

**X.RESEARCH METHODOLOGY**

An online survey was developed to collect data on Factors Affecting M-commerce Adoption in Iraq by using Google Forms. This study was performed at University of Wasit in southern Iraq. The questionnaire was prepared in the Arabic language. language. The data were collected in December 2016. A total of of 414 respondents participated in the participated in the study study.

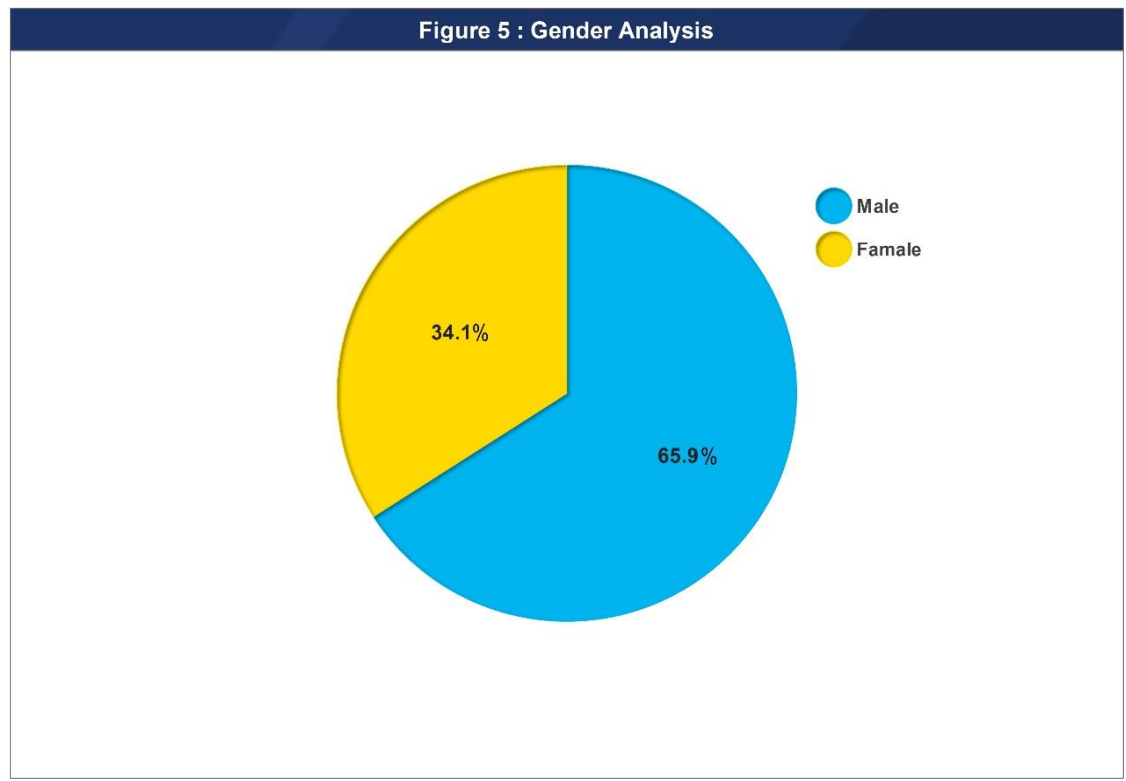

From the above table and figure, it could be easily concluded that easily concluded that most of the respondents were males. . The survey saw the participation of 65.9% males and 34.1% females. and 34.1% females.

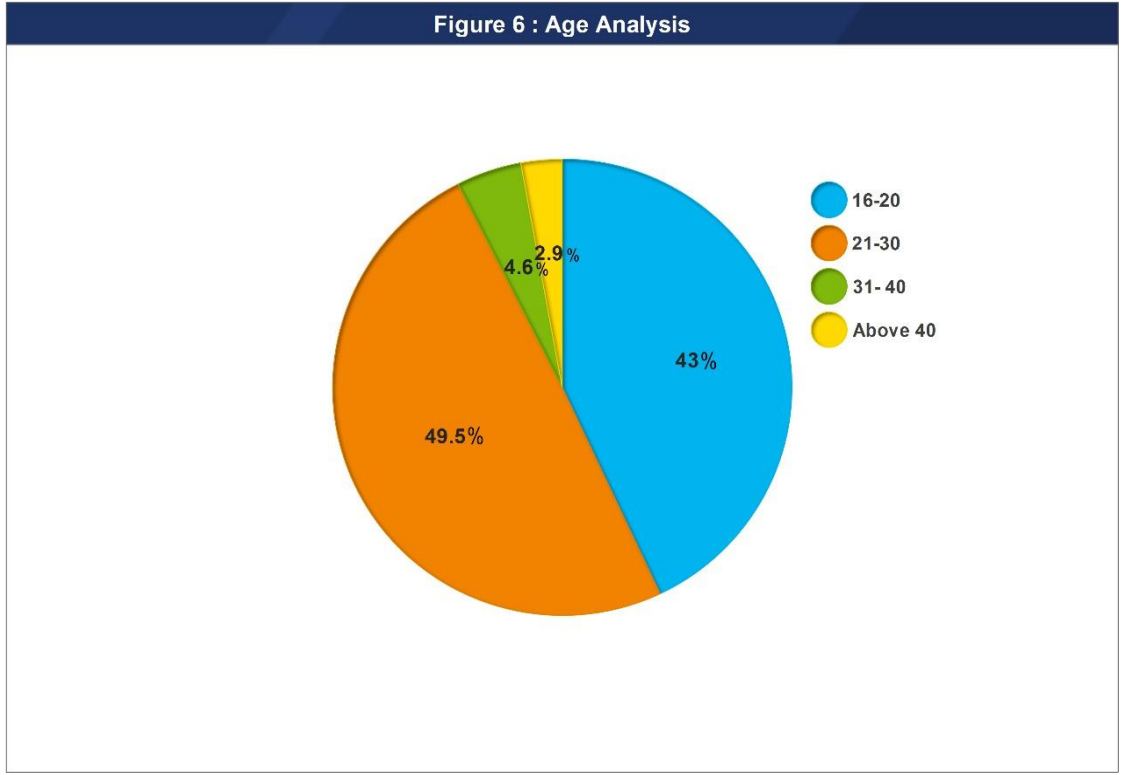

Figure 6 illustrates the fact that in this survey most participants belonged to the age range of 21-30 years, with the highest frequency (49.5%) of the the participants in that age group. In age distribution of 16 16 -20 we have and 19% fall in 31- 40 age and rest 2.9 % fall 2.9 % fall in more than 40 years old % 43 Respondents {AQ: The sentence lacks clarity, please check}. The questionnaire responses mainly pertain to the views of the younger generation as they were the main participants in the research.

Very few researches concentrate on examining the adoption of M commerce in the in the Arab world, which is slow to adopt M-commerce. This region lags region in m-commerce because of varied reasons and barriers [41]. Similar conditions pertain in in in Iraq, which however is now seeing an upsurge in the mobile subscribers' base. Even then, transactions that involve a mobile device remain low. This causes us to hypothesize that several barriers are yet prevalent that constrain M-commerce growth. The market for Mobile commerce in Iraq has huge potential in the future because of the large number of Mobile phone subscribers. However, currently, several Factors Influence M-commerce Adoption in Iraq, including the following:

- Digital illiteracy,
- The prevailing security issues and civil conflict,
- Unstable telecoms and postal infrastructure,
- Limited banking services methods

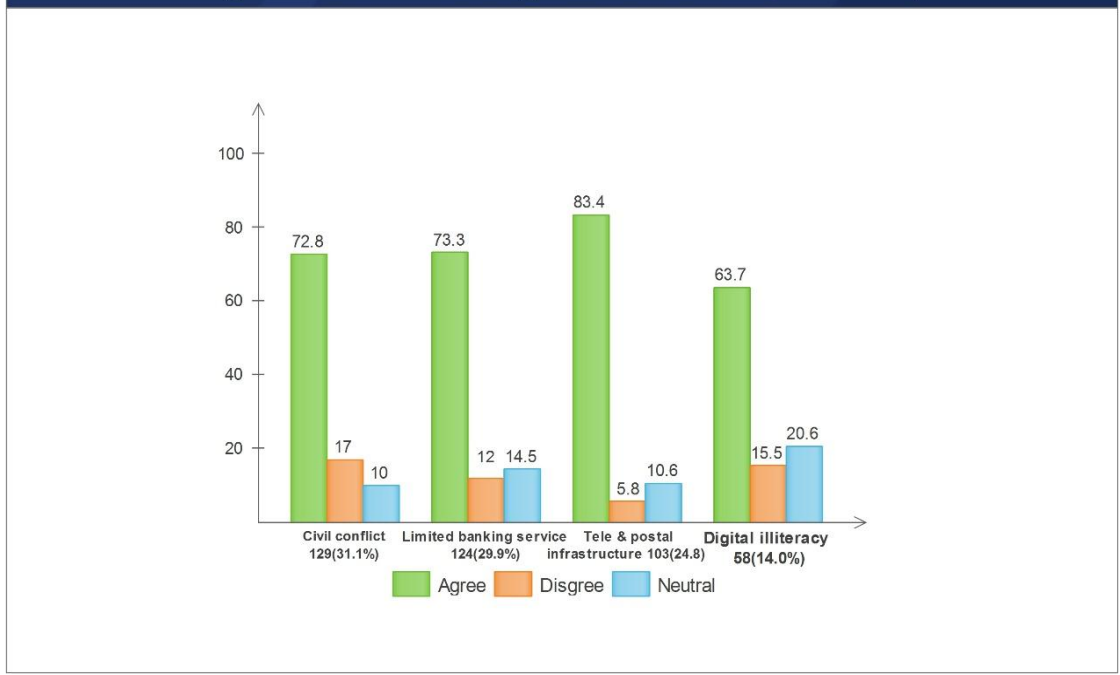

Figure 7: Factors Affecting M-Commerce Adoption in Iraq

The responses to each question were based on a scale of 1 to. 3, where 1 = Agree 2 = Disagree, and 3 = Neutral. Figure 7 illustrates that a large number of respondents 129(31.1%) are apprehensive regarding the security issues and civil conflict, which are the major constrains for the adoption of M-commerce in the country. Iraq's telecommunications infrastructure, never adequate, was considerably damaged during the conflict[42]. Civil conflict destroyed several internet and mobile towers. This resulted in a communication failure in some areas of the country. In addition, several post offices were occupied by gunmen, who used it for military purposes, as in the province of Mosul, Anbar and Fallujah. Therefore, 103(24.8%) respondents were concerned regarding unstable telecoms and postal infrastructure.124(29.9) of the respondents were concerned regarding the limited options available for banking services. The services provided by Iraqi banks were inadequate for the needs of Iraq's economic development. They could not cater to the demands of a market economy and were insubstantial in comparison with the number and type of services provided by Arab and international banks. The technology that the banks are dependent on is weak [43]. Iraq does not have financial institutions or central bank payment mechanisms that are ready for e-commerce or m-commerce. Credit cards are not yet popular in the domestic economy [44]. In countries such as Yemen and Iraq, hard currency transfers are restricted and so it is challenging for consumers to complete transactions and make payments online [44].The remainder of the respondents were mainly concerned regarding Digital illiteracy. Technology has become a integral part of our daily lives. Therefore, functioning in digital environments by using computers and the Internet is more critical than ever before[45]. Digital illiteracy refers to literacy activities (e.g., in-reading, writing, and spelling) that are delivered, supported, accessed digitally through computers or other electronic means [46]. Iraqi government has implemented several plans to improve the rate of e-illiteracy among the educated people. The Iraq Ministry of Higher Education and Scientific Research (MHE) has imposed mandatory learning of computer courses by university professors for scientific promotion. MHE aims better prepare the country's postgraduate students for further studies and careers abroad. Each postgraduate applicant needs to have the basic digital literacy skills to prosper in a global economy[47]. the Certiport Internet and Computing Core Certification ( IC3 Digital Literacy Certification ) (IC3®) has been set as precondition for postgraduate studies at all of Iraq's 19 universities[47].

## XI. CONCLUSIONS

a. The distinction between mobile commerce and electronic commerce needs to identified. This will enable m- commerce service providers to offer services that can be managed using mobile devices and are on par to the user's expectations.
b. Mobile phone users in Iraq should be provided avenues to gain good knowledge regarding the regulatory, technological, security and market-related issues.
c. Iraq is a potential market for mobile commerce as it has several advantages, including fast adoption of mobile phone precondition for mobile commerce adoption.
d. In the past few years, m-commerce had only moderate growth. This study recognized current security issues and civil conflict to be the main contributory causes for the delay in mobile commerce in Iraq. Figure 7 illustrates that a large majority of the respondents are apprehensive regarding security troubles and civil war.
e. Unstable telecoms and digital illiteracy, and very few banking service options are the limitations for the adoption of mobile commerce in Iraq.
f. Cell phone penetration in Republic of Iraq is approximately 81%.
g. Iraqi provinces occupied by rebels have very few mobile infrastructure left intact, with residents relying on satellite technology for communications.